\documentclass{aa}
\usepackage{graphics}

\newcommand{\kms}   {~km~s$^{-1}$}
\newcommand{\mjy}   {~mJy~beam$^{-1}$}

\newcommand{\cmd}   {~cm$^{-2}$}
\newcommand{\cmt}   {~cm$^{-3}$}

\newcommand{\vlsr}  {$v_{\rm LSR}$}

\newcommand{\Tex}   {T_{\rm ex}}
\newcommand{\T}[1]  {T_{\rm #1}}
\newcommand{\hhd}   {HH~2}
\newcommand{\et}    {et al.}
\newcommand{\eg}    {e.\,g.,}
\newcommand{\id}    {{\em id.}}
\newcommand{\nh}    {NH$_3$}
\newcommand{\hco}   {HCO$^+$}
\newcommand{\hcdo}  {HC$^{18}$O$^+$}
\newcommand{\htco}  {H$^{13}$CO$^+$}
\newcommand{\form}  {H$_2$CO}
\newcommand{\cthd}  {C$_3$H$_2$}
\newcommand{\metha} {CH$_3$OH}
\newcommand{\cdo}   {C$^{18}$O}
\newcommand{\tco}   {$^{13}$CO}

\newcommand{\J}[2]  {\mbox{#1--#2}}
\newcommand{\JK}[4] {\mbox{#1$_{#2}$--#3$_{#4}$}}
\newcommand{\Jp}[4] {\mbox{$\frac{#1}{#2},\frac{#3}{#4}$}}
\newcommand{\ap}    {A$^{+}$}

\newcommand{\mm}    {$\pm$}
\newcommand{\mq}    {$\la$}

\newcommand{\N}[1]  {$\, 10^{#1}$}
\newcommand{\ie}    {i.\,e.,}
\newcommand{\arcdeg}{\mbox{$^\circ$}} 
\newcommand{\nodata}{ ~$\cdots$~ }    

\begin{document}


\thesaurus{09(09.09.01 HH 2;09.01.1;09.03.1;09.03.13.2;08.06.02;13.19.3)}

\title{The Molecular Condensations Ahead of Herbig-Haro Objects. I 
Multi-transition Observations of HH~2}

\author{
J. M.      Girart\inst{1}, 
S.           Viti\inst{2},
D. A.    Williams\inst{2},
R.      Estalella\inst{1}, 
and
P. T. P.       Ho\inst{3}
}

\offprints{J.M. Girart, jgirart@am.ub.es}

\institute{
Departament d'Astronomia i Meteorologia, Universitat de Barcelona, 
Av.\ Diagonal 647, 08028 Barcelona, Catalunya, Spain
\and
Department of Physics and Astronomy, University College London,
London, WC1E 6BT, England
\and
Harvard-Smithsonian Center for Astrophysics, 60 Garden Street, Cambridge, 
MA 02138, USA
}

\date{Received ...; accepted ...}

\authorrunning{Girart et al.}
\titlerunning{Multi-transition observations of HH~2}

\maketitle


\begin{abstract}

We present a CSO and BIMA molecular line survey of the dense, quiescent
molecular environment ahead of \hhd.  The molecular gas is cold, 13~K, and
moderately dense, 3\N{5}\cmt.  A total of 14 species has been detected
(including different isotopes and deuterated species). The relative abundances
of the clump are compared with other dense molecular environments, including
quiescent dark clouds, and active low and high mass star forming regions. This
comparison confirms the peculiar chemical composition of the quiescent gas
irradiated by the HH objects. Thus, from this comparison, we found that the
\hco, \metha\ and \form\ are strongly enhanced.  
SO and SO$_2$ are weakly enhanced, whereas HCN and CS are underabundant. 
The CN abundance is within the range of value found in starless dark clouds, 
but it is low with respect to high mass star forming regions. 
Finally, the chemical composition of
\hhd\ confirms the qualitative results of the Viti \& Williams (1999) complex
chemical model that follows the chemical behavior of a molecular clump
irradiated by a HH object.


\end{abstract}

\section{Introduction\label{intro}}

Herbig-Haro (HH) objects, which trace shock-excited plasma, are signposts of 
the intense outflow phenomena associated with star formation (\eg\  
Reipurth \& Bally \cite{Reipurth01}).  Several HH objects are found to have 
associated quiescent dense clumps ahead of them: HH~1/2 (see references below), 
HH~7-11 (Rudolph \& Welch \cite{Rudolph88}; Dent \et\ \cite{Dent93}; Rudolph 
\et\ \cite{Rudolph01}), HH~34 (Rudolph \& Welch~\cite{Rudolph92}), NGC~2264G 
(Girart \et\ \cite{Girart00}), HH~80N (Girart~\et\ \cite{Girart94}, 
\cite{Girart01}; Girart, Estalella \& Ho \cite{Girart98}). Most of them have 
high excitation knots, \ie\ with strong high excitation lines, such as [OIII] 
and H$\alpha$, strong UV lines and significant UV continuum radiation (\eg\ 
Reipurth \& Bally \cite{Reipurth01}).  The quiescent clumps are characterized 
by the low temperature and narrow line widths of the molecular emission, and 
by the emission enhancement of \hco\ and \nh. In spite of the presence of the 
HH objects, there is a lack of apparent dynamical perturbation in the clumps. 
However, recent observations show clear signposts of star formation within the
HH~80N clumps (Girart \et\ \cite{Girart01}).

Girart \et\ (\cite{Girart94}) suggested that the regions of enhanced HCO$^+$
and NH$_3$ emission were a consequence of the irradiation from the HH shock
affecting the chemistry within a small dense clump nearby in the molecular
cloud. It was supposed that the UV radiation would evaporate icy mantles and
promote photochemistry in the enriched gas. Taylor \& Williams
(\cite{Taylor96b}) modeled this situation with a simple chemistry and showed
that HCO$^+$ would arise from the interaction of C$^+$ (derived from CO) with 
H$_2$O liberated from the ice. The chemical enhancements would be transient, 
but predicted column densities were significant. Viti \& Williams 
(\cite{Viti99}, hereafter VW99) noted that the impact of a high radiation 
field on a dense gas should lead to a rich chemistry, and they therefore 
extended the 1996 model.  They predicted that, in addition to HCO$^+$ and 
NH$_3$, a wide variety of species should arise with enhanced abundances, 
including CH$_3$OH, H$_2$CO, SO, SO$_2$, and CN and that the effects in a 
single clump should last for $10^4$ years.  Raga \& Williams (\cite{Raga00}) 
showed that the photochemical effects created by the HH object 
should respond to its movement through the molecular cloud, and predicted that 
the enhanced regions should have a characteristic morphology.

%
     \begin{table*}
     \caption[]{Lines Observed with the 10-m CSO telescope}
     \label{tcso}
     \[
     \begin{tabular}{lcccccc}
     \noalign{\smallskip}
     \hline
     \noalign{\smallskip}     
        &           & $\nu$ & $\!\!\!\!$Beam &$\!\!\int T_{\rm mb} dv$  & \vlsr & $\Delta v^{a}$ \\
Molecule& Transition& (GHz) & $\!\!\!\!$Size &$\!\!\!$(K \kms)&(\kms)&(\kms)   \\
     \noalign{\smallskip}
     \hline
     \noalign{\smallskip}     
CS      & \J{5}{4}	       & 244.93561 & $31''$ & 0.17\mm0.03 &  6.4\mm0.2  &  2.4\mm0.6  \\
\cthd\  & \JK{3}{3,0}{2}{2,1}  & 216.27873 & $34''$ &   $\la$0.07 & \nodata     & \nodata     \\
HCN     & \J{3}{2}	       & 265.88643 & $28''$ & 0.65\mm0.05 & 6.43\mm0.10 &  2.6\mm0.3  \\
\hco\   & \J{3}{2}	       & 267.55763 & $28''$ & 7.53\mm0.08 & 6.84\mm0.01 & 1.83\mm0.03 \\
\hco\   & \J{4}{3}	       & 356.73425 & $21''$ & 3.80\mm0.08 & 6.91\mm0.02 & 1.84\mm0.04 \\
\htco\  & \J{3}{2}	       & 260.25548 & $29''$ & 0.31\mm0.03 & 7.06\mm0.07 & 1.30\mm0.15 \\
HCS$^+$ & \J{5}{4}	       & 213.36053 & $35''$ &   $\la$0.05 & \nodata     & \nodata     \\
H$_2$CO & \JK{3}{0,3}{2}{0,2}  & 218.22219 & $34''$ & 1.23\mm0.04 & 6.73\mm0.03 & 1.77\mm0.07 \\
H$_2$CO & \JK{5}{0,5}{4}{0,4}  & 362.73605 & $21''$ &   $\la$0.12 & \nodata     & \nodata     \\
H$_2$CS & \JK{7}{0,7}{6}{0,6}  & 240.26616 & $31''$ &   $\la$0.08 & \nodata     & \nodata     \\
H$_2$S  & \JK{2}{2,0}{2}{1,1}  & 216.71044 & $34''$ &   $\la$0.07 & \nodata     & \nodata     \\
\metha\ & \JK{5}{0}{4}{0} E    & 241.70017 & $31''$ &   $\la$0.06 & \nodata     & \nodata     \\
\metha\ & \JK{5}{-1}{4}{-1} E  & 241.76722 & $31''$ & 0.21\mm0.03 & 6.54\mm0.06 & 1.24\mm0.14 \\
\metha\ & \JK{5}{0}{4}{0} \ap  & 241.79143 & $31''$ & 0.22\mm0.02 & 6.54\mm0.06 & 1.24\mm0.14 \\
\metha\ & \JK{7}{-1}{6}{-1} E  & 338.34463 & $22''$ &   $\la$0.08 & \nodata     & \nodata     \\
\metha\ & \JK{7}{0}{6}{0} \ap  & 338.40868 & $22''$ &   $\la$0.07 & \nodata     & \nodata     \\
\metha\ & \JK{7}{1}{6}{1} E    & 338.61500 & $22''$ &   $\la$0.08 & \nodata     & \nodata     \\
NS      &\Jp{11}{2}{13}{2}--\Jp{9}{2}{11}{2}$^{b}$
                               & 253.57048 & $30''$ &   $\la$0.06 & \nodata     & \nodata     \\
NS      &\Jp{11}{2}{9}{2}--\Jp{9}{2}{7}{2}
                               & 253.57215 & $30''$ &   $\la$0.06 & \nodata     & \nodata     \\
SO      & \JK{6}{5}{5}{4}      & 219.94939 & $34''$ & 0.70\mm0.03 & 6.52\mm0.02 & 1.26\mm0.06 \\
SO      & \JK{7}{6}{6}{5}      & 261.84370 & $29''$ & 0.31\mm0.03 & 6.62\mm0.04 & 1.16\mm0.13 \\
     \noalign{\smallskip}
     \hline
     \end{tabular}
     \]
     \begin{list}{}{}
     \item{$^{a}$} 
     $\Delta v$ is the FWHM line width
     \item{$^{b}$} Overlapped with the NS 11/2,11/2--9/2,9/2 
     transition
     \end{list}
     \end{table*}
%

Implicit in the model is the idea that molecular clouds are clumpy on a small
scale. This was suggested by Taylor, Morata \& Williams (\cite{Taylor96a}) as a
necessary assumption to interpret the relative spatial distributions of CS and
NH$_3$ in molecular clouds in terms of time-dependent chemistry. The clumps
were required to be transient on a timescale of about one million years. 
Subsequent interferometer observations (e.g. Peng \et\ \cite{Peng98}, for 
TMC-1 Core D; Morata, Girart \& Estalella \cite{Morata02} for L673) have 
supported the Taylor \et\ model.
Therefore, the HH objects may be regarded as probes of the transient
substructure of molecular clouds. Wherever the probe encounters a
small clump, a characteristic chemistry arises. This paper reports a
search for that characteristic chemistry in the clump associated with
\hhd.

The prototypical HH~1 and HH~2 objects (Herbig \cite{Herbig51}; Haro
\cite{Haro52}), located in the L1641N molecular cloud in Orion, provided the
earliest spectral evidence that the HH objects trace strong shocks due to the
presence of stellar winds (Schwartz \cite{Schwartz78}).  Both HH~1 and \hhd\
belong to the class of high excitation HH objects, with strong UV (B\"{o}hm
\et\ \cite{Bohm87}; Raymond, Hartmann \& Hartigan \cite{Raymond88}; Raymond,
Blair \& Long \cite{Raymond97}) and centimeter continuum emission  
(Rodr\'{\i}guez \et\ \cite{Rodriguez90}, \cite{Rodriguez00}). \hhd\ shows a 
very complex and apparently chaotic morphological, kinematical and excitation 
structure (\eg\ Schwartz \et\ \cite{Schwartz93}; Eisl\"{o}ffel \et\ 
\cite{Eisloffel94}).  Detailed HST observations reveal that this complex 
structure is the consequence of the interaction of the \hhd\ jet with a dense 
ambient cloud, resulting in a very bright, high ionization and complex 
structure (Hester, Stapelfeldt, \& Scowen \cite{Hester98}).  Recent Chandra 
observations reveal that the strongest \hhd\ knot, which is also the highest 
excitation knot, has associated X-ray emission (Pravdo \et\ \cite{Pravdo01}).

The presence of quiescent dense ambient gas ahead of \hhd\ has been well
established from \hco\ and \nh\ observations (Davis, Dent \& Bell Burnell
\cite{Davis90}; Torrelles \et\ \cite{Chema92}, \cite{Chema94}; Choi \& Zhou
\cite{Choi97}).  The ammonia emission shows a clumpy medium, with clump sizes
of only 20$''$ (Torrelles \et\ \cite{Chema92}) or $\sim 8000$~AU (assuming a
distance of 390~pc: Anthony-Twarog \cite{Anthony82}).  JCMT observations by
Dent (\cite{Dent97}) shows that the chemical composition of the dense ambient 
gas is strongly altered by the presence of the \hhd\ radiation, with a strong 
emission enhancement of the \hco. Wolfire \& K\"{o}nigl (\cite{Wolfire93}) 
carried out a shock model where the X-ray, EUV and FUV radiation field from 
\hhd\ penetrates the cloud, initiates an ion chemistry, enhancing the 
abundances of the \hco\ and the electrons, which produces an excitation 
enhancement of the \hco. The combination of these two effects lead to a 
strong emission enhancement.

In this paper we present 0.8--1.4~mm CSO observations towards the quiescent
clump ahead of \hhd. We also present 3~mm spectra   obtained with the BIMA
array.  The goal of this paper is to study the chemical  effect produced by the
strong \hhd\ radiation over the dense quiescent ambient  gas around \hhd.   A
second paper will analyze and discuss the spatial  distribution of the
molecules observed with the BIMA array (Girart \et\ \cite{GEHVW02}, hereafter
Paper II).

\section{Observations and Results}

     \begin{table*}
     \caption[]{Lines Observed with the BIMA Array}
     \label{tbima}
     \[
\begin{tabular}{lcrccc}
     \noalign{\smallskip}
     \hline
     \noalign{\smallskip}   
\multicolumn{2}{c}{} &
\multicolumn{1}{c}{$\nu$} &
\multicolumn{1}{c}{$\!\!\int T_{\rm mb} dv$} &
\multicolumn{1}{c}{\vlsr} & 
\multicolumn{1}{c}{$\Delta v^a$} 
\\
\multicolumn{1}{l}{Molecule} &
\multicolumn{1}{c}{$\!\!$Transition} &
\multicolumn{1}{c}{(GHz)} &
\multicolumn{1}{c}{$\!\!\!\!$(K \kms)} &
\multicolumn{1}{c}{(\kms)} &
\multicolumn{1}{c}{(\kms)} 
\\
     \noalign{\smallskip}
     \hline
     \noalign{\smallskip}     
CH$_3$CN &\JK{6}{1}{5}{1} F=7-6&110.38140 &   $\la$0.10 & \nodata     & \nodata     \\
CH$_3$CN &\JK{6}{0}{5}{0} F=7-6&110.38352 &   $\la$0.10 & \nodata     & \nodata     \\
CH$_3$SH & \JK{3}{0}{2}{0} \ap & 75.86293 &   $\la$0.08 & \nodata     & \nodata     \\
\metha\  & \JK{2}{-1}{1}{-1} E & 96.73939 & 0.33\mm0.03 & 6.57\mm0.03 & 0.86\mm0.04 \\
\metha\  & \JK{2}{0}{1}{0} \ap & 96.74143 & 0.47\mm0.03 & \id         & \id         \\
\metha\  & \JK{2}{0}{1}{0} E   & 96.74458 & 0.06\mm0.03 & \id         & \id         \\
\metha\  & \JK{2}{1}{1}{1} E   & 96.75551 &   $\la$0.07 & \id         & \id         \\
CN      & $\!\!\!\!\!$ \J{1}{0} \Jp{1}{2}{3}{2}--\Jp{1}{2}{3}{2} 
                               &113.19132 & 0.09\mm0.01 & 7.10\mm0.09 &  1.2\mm0.2  \\
CN      & $\!\!\!\!\!$ \J{1}{0} \Jp{3}{2}{3}{2}--\Jp{1}{2}{1}{2} 
                               & 113.48814 &  $\la$0.08 & \nodata     & \nodata     \\
CN      & $\!\!\!\!\!$ \J{1}{0} \Jp{3}{2}{5}{2}--\Jp{1}{2}{3}{2} 
                               &113.49098 & 0.17\mm0.01& 6.77\mm0.07  &  1.5\mm0.2 \\
CN      & $\!\!\!\!\!$ \J{1}{0} \Jp{3}{2}{1}{2}--\Jp{1}{2}{1}{2} 
                               &113.49964 & 0.12\mm0.01 & \id         & \id         \\
\cdo\   & \J{1}{0}             &109.78216 & 0.38\mm0.02 & 6.43\mm0.01 & 0.53\mm0.03 \\
\tco\   & \J{1}{0}             &110.20135 & 4.38\mm0.05 & 7.09\mm0.02 & 1.76\mm0.02 \\
CS      & \J{2}{1}	       & 97.98097 & 0.55\mm0.04 & 6.51\mm0.06 & 1.72\mm0.12 \\
\cthd\  & \JK{2}{1,2}{1}{0,1}  & 85.33890 &   $\la$0.06 & \nodata     & \nodata     \\
DCN     & \J{1}{0}\ F=2--1     & 72.41491 & 0.51\mm0.03 & 6.78\mm0.04 & 1.17\mm0.08 \\
DCO$^+$ & \J{1}{0}\            & 72.03933 & 0.74\mm0.06 & 6.59\mm0.03 & 0.76\mm0.07 \\
\hcdo\  & \J{1}{0}             & 85.16226 & 0.07\mm0.02 &  6.9\mm0.2  &  1.6\mm0.3  \\
HCN     & \J{1}{0}\ F=1--1     & 88.63042 & 0.60\mm0.04 & 6.44\mm0.03 & 1.97\mm0.04 \\
HCN     & \J{1}{0}\ F=2--1     & 88.63185 & 1.40\mm0.05 & \id         & \id         \\
HCN     & \J{1}{0}\ F=0--1     & 88.63394 & 0.36\mm0.04 & \id         & \id         \\
\hco\   & \J{1}{0}	       & 89.18852 & 8.48\mm0.05 & 6.83\mm0.01 & 1.73\mm0.01 \\
HCOOH   & \JK{4}{0,4}{3}{0,3}  & 89.57917 &   $\la$0.07 & \nodata     & \nodata     \\
HCS$^+$ & \J{2}{1}	       & 85.34790 &   $\la$0.06 & \nodata     & \nodata     \\
HNCO    & \JK{5}{0,5}{4}{0,4}  &109.90576 &   $\la$0.08 & \nodata     & \nodata     \\
H$_2$CO & \JK{1}{0,1}{0}{0,0}  & 72.83795 & 2.46\mm0.09 & 6.54\mm0.02 & 1.15\mm0.05 \\
H$_2$CO & \JK{5}{1,4}{5}{1,5}  & 72.40909 &   $\la$0.07 & \nodata     & \nodata     \\
H$_2$CS & \JK{3}{1,3}{2}{1,2}  &101.47775 &   $\la$0.06 & \nodata     & \nodata     \\
H$_2$CS & \JK{3}{0,3}{2}{0,2}  &103.04040 &   $\la$0.07 & \nodata     & \nodata     \\
H$_2$CS & \JK{3}{2,1}{2}{2,0}  &103.05179 &   $\la$0.07 & \nodata     & \nodata     \\
HC$_3$N & \J{11}{10}	       &100.07639 &   $\la$0.05 & \nodata     & \nodata     \\
NH$_2$CN& \JK{5}{1,4}{4}{1,3}  &100.62950 &   $\la$0.03 & \nodata     & \nodata     \\
OCS     & \J{8}{7}             & 97.30121 &   $\la$0.06 & \nodata     & \nodata     \\
SO      & \JK{2}{2}{1}{1}      & 86.09399 & 0.44\mm0.04 & 6.47\mm0.04 & 0.79\mm0.11 \\
SO      & \JK{3}{2}{2}{1}      & 99.29988 & 1.77\mm0.04 & 6.45\mm0.03 & 1.01\mm0.03 \\
SO      & \JK{4}{5}{4}{4}      &100.02957 &   $\la$0.05 & \nodata     & \nodata     \\
SO      & \JK{2}{3}{1}{2}      &109.25218 & 0.35\mm0.04 & 6.60\mm0.03 & 0.89\mm0.09 \\
SO$_2$  & \JK{6}{0,6}{5}{1,5}  & 72.75824 & 0.26\mm0.05 &  6.8\mm0.2  &  1.3\mm0.3  \\
SO$_2$  & \JK{10}{1,9}{9}{2,8} & 76.41217 &   $\la$0.21 & \nodata     & \nodata     \\
SO$_2$  & \JK{3}{1,3}{2}{0,2}  &104.02942 & 0.22\mm0.01 & 6.59\mm0.02 & 0.97\mm0.04 \\
SO$_2$  &\JK{10}{1,9}{10}{0,10}&104.23929 &   $\la$0.04 & \nodata     & \nodata     \\
     \noalign{\smallskip}
     \hline
     \end{tabular}
     \]
     \begin{list}{}{}
     \item{$^{a}$} $\Delta v$ is the FWHM line width
     \end{list}
     \end{table*}
%

%
     \begin{figure*} 
     \resizebox{\hsize}{!}{\includegraphics{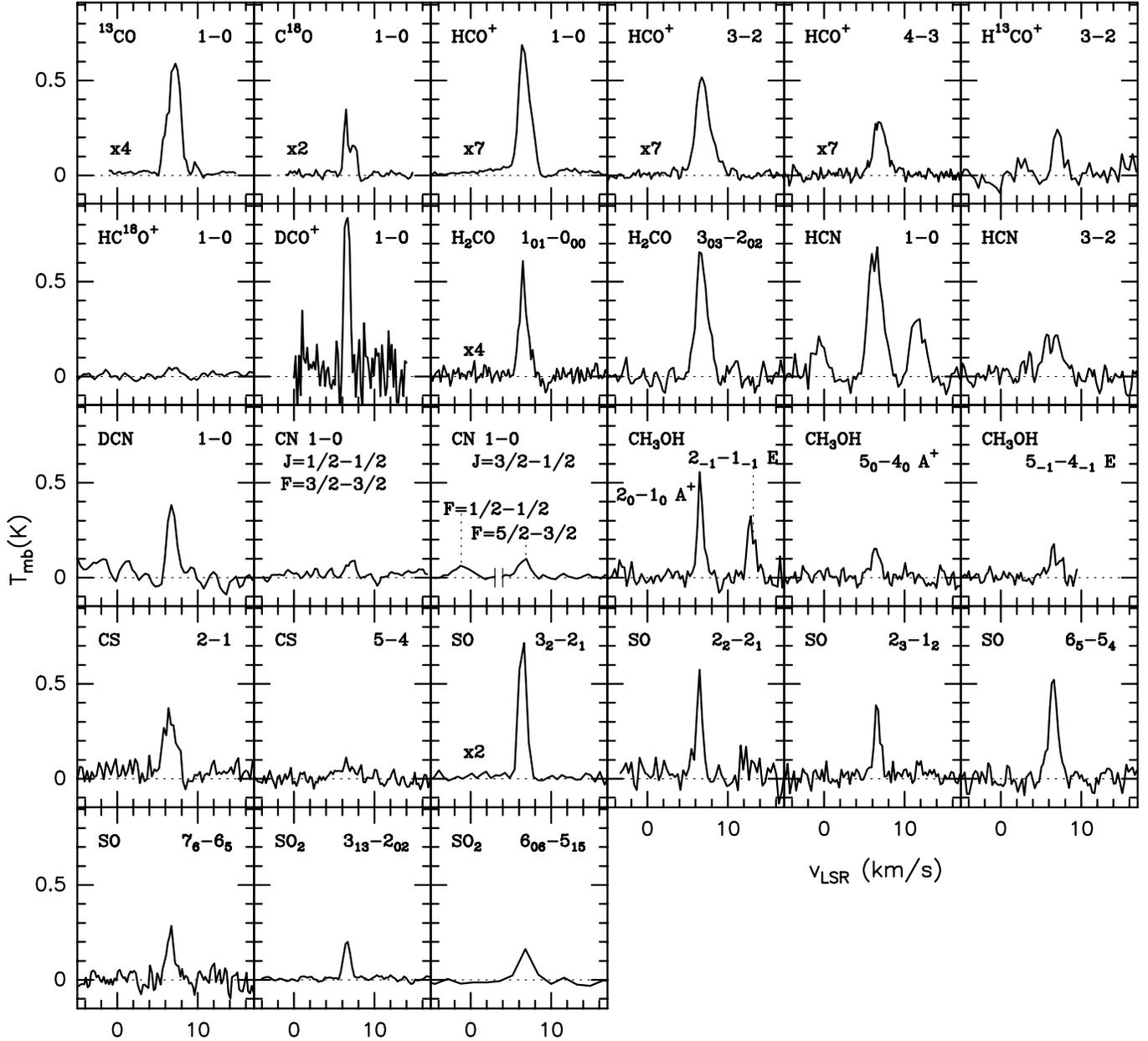}}
     \caption[]{
Spectra of the detected lines with the BIMA and CSO telescopes. Lines from BIMA
were convolved with a Gaussian in order to obtain a resulting beam of 30$''$,
approximately the beam of the CSO data at 250~GHz.
}
     \label{fspec} 
     \end{figure*}
%

Observations were carried out in October 1999 with the 10.4~m telescope of the
Caltech Submillimeter Observatory\footnote{The Caltech  Submillimeter
Observatory is funded by the National Science Foundation under  contract
AST-9615025} (CSO).   The 200-300 GHz and 300-400~GHz receivers were used in
conjunction with  50 MHz and 500 MHz acousto-optical spectrometers, which
provided velocity  coverage of 60 and 600~\kms\ at 250~GHz and of 40 and
400~\kms\ at 350~GHz.   Pointing, focus and calibration were checked every few
hours using Saturn. The main-beam efficiency, measured by observing Saturn,
were $\sim$67\% at  250~GHz and $\sim$60\% at 350~GHz and system temperatures
ranged 240--670~K.  All the spectra were observed at the position:  
$\alpha (J2000) = 5^{\rm h}36^{\rm m}26\fs84$;  
$\delta (J2000) = -6\arcdeg47'27$\farcs$1$. 
This position is very close ($\sim 3''$) to the \hhd\ knot L (\eg\ Hester \et\
\cite{Hester98}) and to the peak emission of the \hco\  \J{4}{3}\ integrated
map from Dent (\cite{Dent97}). 
Table~\ref{tcso} lists all the molecular lines observed along with their rest
frequency, the angular resolution at the observed frequency and the Gaussian
fits to the line profiles.  The upper limits show the 3-$\sigma$ level within
the  5.2-7.8~\kms\ range.

The 10--antenna BIMA array\footnote{ The BIMA array is operated by the
Berkeley--Illinois--Maryland Association  with support from the National
Science Foundation.} observations were carried out between 1999 October and 
2001 May in the C configuration.  A detailed description of the  observations
and reduction is given in Paper II. In order to compare with the  CSO spectra,
we smoothed the BIMA maps by convolving them with a Gaussian,  with a resulting
FWHM beam of $30''$, the angular resolution of the CSO  observations at
230~GHz.  A spectrum of each transition observed with BIMA  was synthesized  at
the position observed with the CSO telescope (see paragraph  above). 
Table~\ref{tbima} lists all the molecular lines observed, their rest 
frequency, and the Gaussian fits to the line profiles. 

Figure~\ref{fspec} shows the spectra of the detected lines with the CSO and 
the BIMA observations smoothed to the $30''$ angular resolution. A total of 33
transitions (including hyperfine transitions) were detected for 14 species:
\tco\ and \cdo, \hco, \htco, \hcdo, HCN, \form, CN, \metha, CS, SO, SO$_2$ and
the deuterated  species DCO$^+$ and DCN. The spectral lines peak at a \vlsr\
within the range of $\sim 6.4$~\kms\ (CS and HCN) to $\sim 6.9$~\kms\ (\hco).
This range of values is in agreement with previous observations of the 
quiescent gas ahead of \hhd\ (Torrelles \et\ \cite{Chema92}; Choi \& Zhou 
\cite{Choi97}). The line widths range from $\sim 0.8$~\kms\ for the DCO$^+$ to 
$\sim 1.9$~\kms\ for \hco\ and HCN. The 1~mm CS and HCN transitions have large
line widths, $\sim$2.4~\kms.  However, this value is not significant due to the
low SNR of these lines.  The differences in line center velocity and line width
are significant and are due to the combination of velocity gradients within the
quiescent ambient gas and a spatial chemical differentiation, which will be
discussed in Paper II.  Yet, for the \hco\ the larger line width can also be due
to a high optical  depth and the presence of high velocity emission.  The \hco\
\J{1}{0}\ spectrum shows clearly redshifted and blueshifted wings, which 
implies the presence of high velocity molecular gas. This high velocity 
component is tracing the interaction between the HH object and the clump,
which  confirms the spatial association of the clump with \hhd\ (Paper II).

\subsection{HH2 and the clump}

%
     \begin{figure} 
     \resizebox{\hsize}{!}{\includegraphics{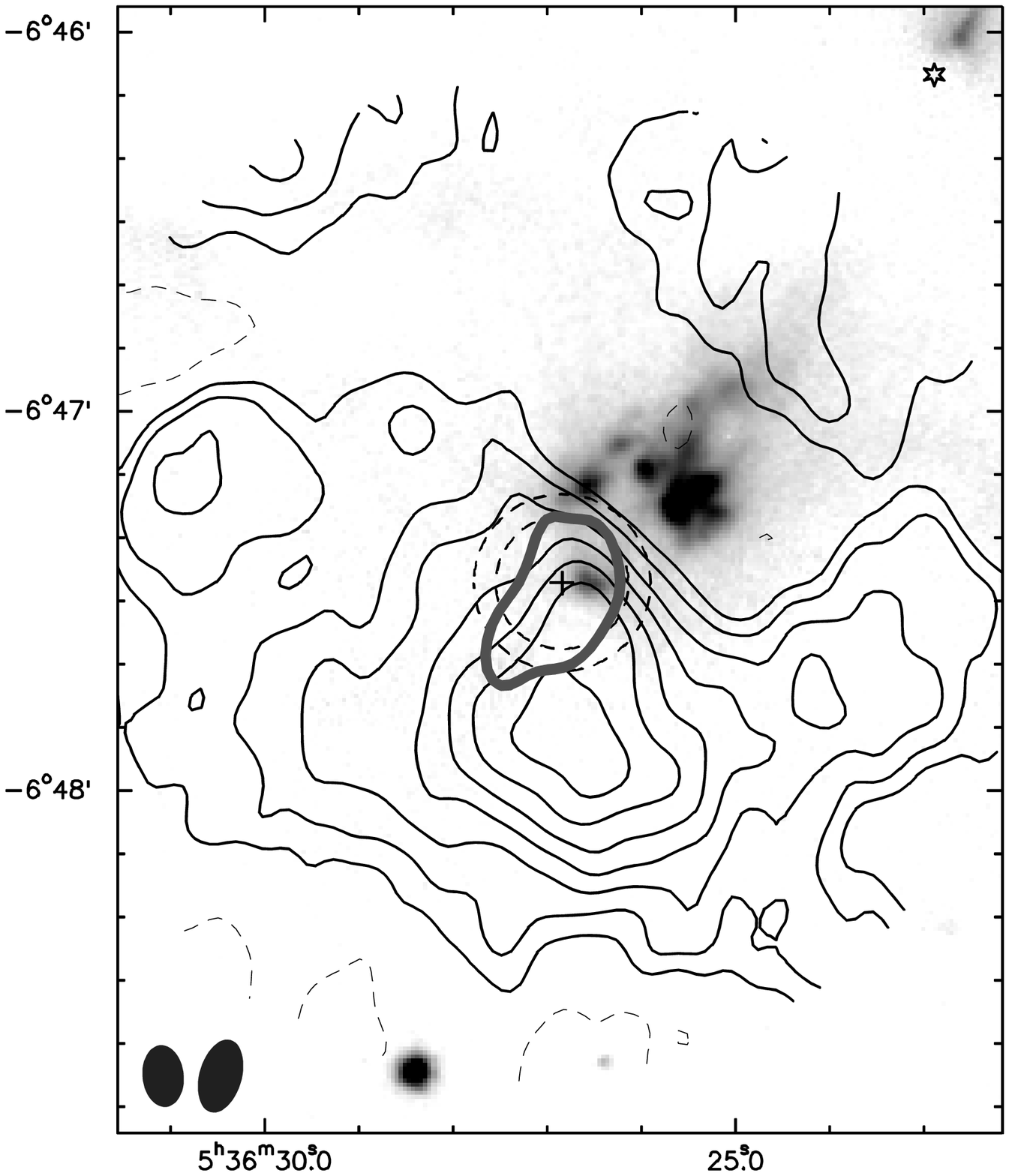}}
     \caption[]{
Superposition of the gray scale of the [SII] image (from Curiel, private
communication), the integrated emission of the \tco\ \J{1}{0}\ line (thin
contours) over the 5.1-8.1\kms\ \vlsr\ interval, and the integrated emission 
of the SO \JK{3}{2}{2}{1}\ line (thick contour) over the aforementioned 
\vlsr\ range.  Contours levels are 4, 13, 25, 37, 49, ... times the rms noise
of the map, 47\mjy\ for the \tco, and 50\% of the peak intensity for the
SO. The synthesized beam of the \tco\ (left) and the SO (right) are shown in 
the lower left corner.  
The star marks the position of \hhd\ VLA~1, the 
powering source of the HH~1/2 outflow.  
The cross marks the center coordinates of the 
single--dish spectra. The two concentric dashed circles show the 21$''$ and
28$''$ beam sizes of the CSO spectra at 265 and 360~GHz, respectively.
}
     \label{fmap} 
     \end{figure}
%

Figure~\ref{fmap} shows the distribution of the quiescent molecular gas, traced
by the \tco\ emission, ahead of the \hhd\ region.  There is clear column density
enhancement, mainly downstream of HH2, whereas at the position of \hhd\ or just
behind it there is little ambient gas. The SO emission is also shown to better
locate the chemically enhanced region just ahead of \hhd, almost coincident with
\hhd\ knot L (\eg\ Hester \et\ \cite{Hester98}).  The high excitation \hhd\
emission, as well as the X-ray and the radio continuum emission, arise from 
the brightest and largest knot in the [SII] image (see Fig~\ref{fmap}).

\section{Analysis}

\subsection{The Physical Conditions of the Quiescent Gas\label{physical}}

\hco, SO and \metha\ are the molecules with more than three transitions 
detected (including the rarer isotopes for the \hco).  Therefore, the observed
emission from these molecules can be used to derive the physical condition of
the quiescent gas. SO and \metha\ are molecules with moderate dipole moments,
1.55 and 0.90~Debyes, respectively, which implies that if the densities are not
too low their emission will be close to thermalization.  Thus, we carried out
the population diagram analysis (\eg\ Goldsmith \& Langer \cite{Goldsmith99})
to derive the temperature of the gas and the column densities of SO and 
\metha. For the \hco, since we have two rare isotopes observed, \htco\ and 
\hcdo, we carried out the Monte Carlo radiative transfer model developed by 
Hogerheijde \& van der Tak (\cite{Hogerheijde00}).

\subsubsection{Population Diagram Analysis\label{pop}}

\begin{figure} 
     \resizebox{\hsize}{!}{\includegraphics{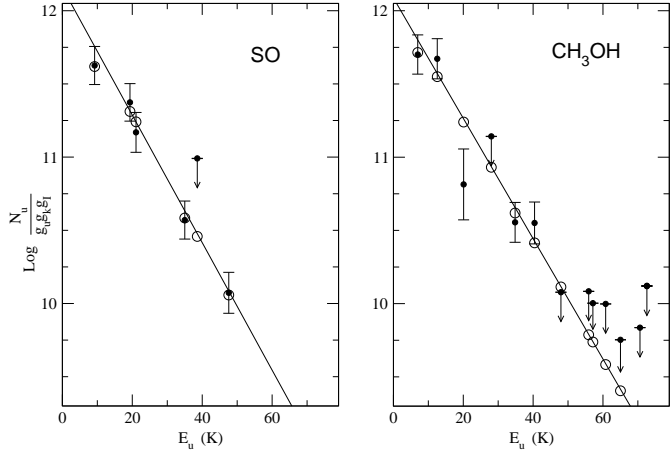}}
     \caption[]{
Population diagrams for the SO (left panel) and the \metha\ (right panel).
Solid filled circles indicate the observed data. The error bars include the
measurement uncertainty and the calibration uncertainty ($\sim 20$\%).  The
solid lines show the best fit obtained by assuming optically thin
emission (\ie\ using the standard population diagram analysis).  The open 
circles shows the best solution for the least square fit taking into account 
the optical depth correction (see \S~\ref{Apendix}).
}
     \label{frot} 
     \end{figure}

The method used to carry out the population diagram analysis, which takes into
account approximately the optical depth, is described in  appendix
\S~\ref{Apendix}.  In order to obtain more accurate fits for the SO and
\metha, which were observed with two different telescopes, a 20\% calibration
uncertainty was included in the estimated errors.  Figure~\ref{frot} shows the
population diagram for the SO and \metha\  molecules.  

%
%

The SO emission is best fitted for $T_{\rm rot}\simeq9.4\pm1.3$~K,
$N$(SO)$\simeq3.2^{+1.4}_{-1.2}\times10^{13}$\cmd\ and a filling factor of
$f=0.35$.  This solution gives moderate optical depths for the observed
transitions, with a maximum value of 1.0 for the  \JK{3}{2}{2}{1}\ transition. 
The high angular resolution ($\theta_{\rm FWHM} \sim 9''$) BIMA SO
\JK{3}{2}{2}{1}\ maps show that the integrated emission arises from a region
with a deconvolved size of $26''\times15'', \, PA=-34\arcdeg$ (Paper II), which
implies a filling factor of approximately 
$f_{\rm SO } \simeq (26\times15)/(26\times15+30\times30)=0.30$.  
Therefore, the best solution for the SO
obtained using the method described in appendix \S~\ref{Apendix}, gives a
filling factor in agreement with the observed value from the BIMA maps.   

The population diagram technique for the \metha\ emission gives the  best fit
for $T_{\rm rot}\simeq10.2\pm0.9$~K,
$N$(\metha)$\simeq6.4^{+1.4}_{-1.0}\times10^{13}$\cmd, and $f=0.20$.  The
optical depths of the observed transitions are small: the maximum value is
0.56, for the \JK{2}{0}{1}{0}\ \ap\ transition. The smaller filling factor of
the \metha\ solution with respect to the SO is because of its slightly smaller
emitting area: $f_{\rm CH_3OH} \simeq 0.24$ (Paper II). We note that although
the optical depths of the SO and \metha\ are small ($\la 1$) the population
diagram technique used here and described in \S~\ref{Apendix} gives column
densities of 20\% (\metha) to 30\% (SO) higher than the standard population
diagram technique where optically thin emission is assumed.

\subsubsection{Monte Carlo Radiative Transfer Code\label{mcarlo}}

We used the one dimensional version of the Monte Carlo code developed by
Hogerheijde \& van der Tak (\cite{Hogerheijde00}), which calculates the
radiative transfer and excitation of molecular lines. The code is formulated
from the viewpoint of cells rather than photons, which allows the separation of
local and external contributions of the radiation field.  This gives an
accurate and fast performance even for high opacities (for more details see
Hogerheijde \& van der Tak  \cite{Hogerheijde00}).  

%
     \begin{figure} 
     \resizebox{\hsize}{!}{\includegraphics{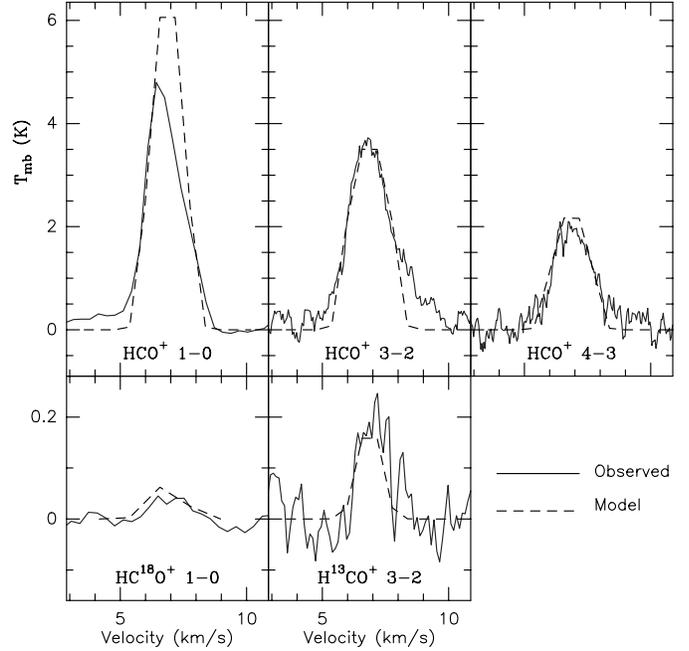}}
     \caption[]{
Observed (solid contours) and synthetic (dashed contours) \hco, \htco\ and
\hcdo\ spectra.  
}
     \label{fhcop} 
     \end{figure}
%

The \hco\ emission was assumed to arise from a sphere of radius $1 \,
10^{17}$~cm or 6680~AU (17$''$ at the distance of \hhd), which is approximately
the radius of the BIMA \hco\ \J{1}{0}\ emission for the peak intensity
component (the \hco\ emission extends over a larger region to the NE of the
peak intensity, but this extended component is not picked up by the CSO beam). 
The volume density and temperature were assumed constant and no velocity
gradient was adopted (we note that the results are insensitive for radial
velocity gradients of $\sim 0.3$\kms\ within the adopted radius, \ie\
$\sim$9\kms\ pc$^{-1}$).  The $^{12}$C$^{16}$O to $^{13}$C$^{16}$O and
$^{12}$C$^{16}$O to $^{12}$C$^{18}$O abundance ratios adopted were 63 and 560,
derived for the Orion region by  Langer \& Penzias  (\cite{Langer90},
\cite{Langer93}).  We explored a large range of values for the volume density,
the temperature, the molecular fractional abundance and the intrinsic line
width. The resulting synthetic maps of the different transitions were convolved
with a Gaussian to match the angular resolution of the spectra ($28''$ for the
\J{1}{0}\ and \J{3}{2}\ transitions and $21''$ for the \J{4}{3}\ transition). 
The best fit is obtained for a volume density of $\sim 2.8 \, 10^{5}$\cmt, a
temperature of $\sim 13$~K, an intrinsic line width of $\sim0.6$\kms, and a
\hco\ fractional abundance of $2.0 \, 10^{-9}$.  Figure~\ref{fhcop} shows the
overlap of the observed and synthetic spectra for the \hco, \htco\ and \hcdo\
transitions.  The synthetic spectra fit well the observed spectra, except for
the \hco\ \J{1}{0}.  The lack of observed emission at the central and
redshifted sections of the line, as compared with the predicted spectra, is
possibly due to a combination of two effects.  First, it could be due to the
missing flux, a consequence of the lack of short spacing  visibilities at the
interferometer.  Second, it is possible that the \hco\  \J{1}{0}\ is affected
by the absorption of a cold and low density component of  the cloud, whose
effects will only be observable in the lowest rotational  transition of
abundant species with high dipole moment (see Girart \et\  \cite{Girart00}).

The assumed radius, derived from the \hco\ \J{1}{0}\ maps, could be highly
uncertain due to high optical depth, absorption by the foreground cold gas or
resolving out the large-scale \hco\ emission.  However, the models run 
with radii higher or lower than roughly a 30\% of the value assumed do not fit
the observed spectra. For small changes in the radii the best solutions are
always in the  12-14K temperature range. 

The temperature derived here is only slightly higher than the temperature
derived from the SO and \metha\ population diagrams. This implies that the
observed SO and \metha\ transitions are close to thermalization for the derived
volume density, $2.8\times10^{5}$\cmt.

\subsection{Excitation Temperature and Column Densities}

From the previous analyses the derived temperature of the cloud is $\sim 13$~K.
Yet, the rotational transitions of the observed molecules may have excitation
temperatures significantly lower than this value because of subthermalization,
\ie\ the critical density of the transition is higher than the volume density of
the cloud, $2.8\times10^{5}$\cmt.  In particular this is the case for the high
dipole moment molecules like \hco\ (3.90 Debyes), HCN (2.98 Debyes), H$_2$CO
(2.33 Debyes) or CS (1.98 Debyes). Thus, for example, the critical density of
the H$_2$CO \JK{3}{0,3}{2}{0,2}\ and CS \J{5}{4}\ are 6\N{6} and 7\N{6} \cmt,
respectively (Choi \& Zhou \cite{Choi97}).

If at least two transitions are detected and the emission is optically thin, 
the excitation temperature can be obtained from the following two expressions:
\begin{equation}\label{eA}
\frac{T_i}{T_j} \, =  \,
\frac{f_i}{f_j} 
\frac{J_{\nu_i}(\Tex)-J_{\nu_i}(\T{bg})}{J_{\nu_j}(\Tex)-J_{\nu_j}(\T{bg})} 
\frac{1-e^{-\tau_i}}{1-e^{-\tau_j}} 
\end{equation}
and
\begin{equation}\label{eB}
\frac{\tau_i}{\tau_j} \, =  \,
\frac{g_{{\rm K}i} \, g_{{\rm I}i}}{g_{{\rm K}j} \, g_{{\rm I}j}} 
\frac{e^{h \nu_i / k \Tex}-1}{e^{h \nu_j / k \Tex}-1} 
e^{-(Eu_j-Eu_i) / k \Tex}
\end{equation}
where the subindices $i$ and $j$ refer to the rotational transitions, $f$ is
the filling factor, $J_{\nu}{(\rm T})$ is the Planck function in temperature
units, $g_{\rm K}$ and $g_{\rm I}$ are the K-level and nuclear spin
degeneracies, respectively, $Eu$ is the energy of the upper level of the
transition, and $\tau$ the optical depth. In most of the cases here, the
spectra of the transitions used to derive $\Tex$ were obtained at the same
angular resolution, so $f_i/f_j=1$.  \hco, and CO are the only molecules with
rarer isotopes observed.  Therefore, for the other molecules, their emission 
is assumed to be optically thin. 

The beam-averaged column densities of the observed species were derived
assuming that the excitation temperature, $\Tex$, is the same for all the
rotational transitions of the same species, with the exception of the SO and
\metha, whose column densities were derived from the Population Diagram
(\S~\ref{pop}).

\subsubsection{\hco}

The \hco\ column density can be derived from the model carried out using the
Monte Carlo radiative transfer code (\S~\ref{mcarlo}) or directly from the line
intensity ratios as described earlier (Eq.~\ref{eB} and \ref{eA}). From the
model  derived in \S~\ref{mcarlo}, the \hco\ beam-averaged column density for
the best fitted model is 1.0\N{14}\cmd. On the other hand, for the intensity
ratio method, the filling factor ratio of the \J{4}{3}\ and \J{3}{2}
transitions, $f_{21''}/f_{28''}$, should be derived in order to properly use
Eq.~\ref{eB} and \ref{eA}.  From the FWHM of the \hco\ \J{1}{0}\ emission,
34$''$ (see \S~\ref{mcarlo}) we can obtain a rough estimation of this ratio:  
$f_{21''}/f_{28''} \simeq (21^2 + 34^2)/(28^2 + 34^2) = 0.8$.  
Thus from the line ratio analysis we estimate
that the \htco\ \J{3}{2}\ transition is optically thin ($\tau < 1$) and that
the excitation temperature is within the 6.2-7.0~K range. Therefore, from the
\htco\ \J{3}{2}\ integrated line intensity, the \hco\ beam-averaged column
density is within the 3-5$\times 10^{13}$\cmd\ range.  The line intensity ratio
analysis derives an excitation temperature lower than the kinetic temperature
obtained in \S~\ref{mcarlo} and also a lower column density (by a factor 2-3).
Uncertainties in the $f_{21''}/f_{28''}$ derived from the \hco\ \J{1}{0}\ maps
(see \S~\ref{mcarlo}) could account for this discrepancy.
Alternatively, non-LTE effects, due to subthermalization, could also cause
this discrepancy.

Since the DCO$^+$ has a similar dipole moment as the \hco, we adopted the
same excitation temperature derived from the line ratios of the \hco: within
the 6-7~K range, the column density of the DCO$^+$ column density is
$1.0\times10^{12}$\cmd.  This value is two orders of magnitude lower than that
of the \hco.

\subsubsection{HCN}
 
For the HCN we fitted the three hyperfine components of the \J{1}{0}\
transitions using CLASS. The best solutions of the fit give moderate or low
optical depths, $\tau \la 1.0$, although the excitation temperature is not well
constrained, $\Tex \ga 4$~K.   However, the observed relative intensities of
the F=1--1, 2--1 and 0--1 transitions, 1.7\mm0.2:3.9\mm0.4:1.0, do not agree
with the 3:5:1 value expected under LTE conditions, but it is within the range
of values measured  in other molecular clouds (Guilloteau \& Baudry
\cite{Guilloteau81}; Wannier \et\ \cite{Wannier74};  Walmsley \et\
\cite{Walmsley82}).  Velocity gradients and a non-uniform  density could cause
this ``anomaly'' (Guilloteau \& Baudry \cite{Guilloteau81}; Cao \et\
\cite{Cao93}; Gonzalez--Alfonso \& Cernicharo \cite{Gonzalez93}).   The
\J{3}{2}\ to \J{1}{0}\ integrated intensity ratio yields an excitation
temperature of $\sim 6$~K, significantly lower than that derived from the
population diagram of the SO and \metha.  This implies that the line emission
is subthermalized, which could also cause the ``anomaly'' in the \J{1}{0}\
hyperfine line ratios. The HCN beam-averaged column density was derived using
$\Tex = 6$~K.

For the DCN, we adopted the same excitation temperature as the one for
the HCN, 6~K.

\subsubsection{\form}

The two transitions detected yield an excitation temperature of $\sim 6$~K,
implying that the emission is subthermalized (the upper limits of the
undetected transitions are not tight enough to better constrain this value). As
in the case of HCN and \hco\, this is due to its high dipole moment. In order
to calculate the \form\ beam-averaged column density we used this excitation
temperature and adopted an ortho to para ratio of 1.5, which should be
reasonable for the \form\ formation and/or ortho-para conversion on grain
surfaces (Dickens \& Irvine \cite{Dickens99}).

%
     \begin{figure} 
     \resizebox{\hsize}{!}{\includegraphics{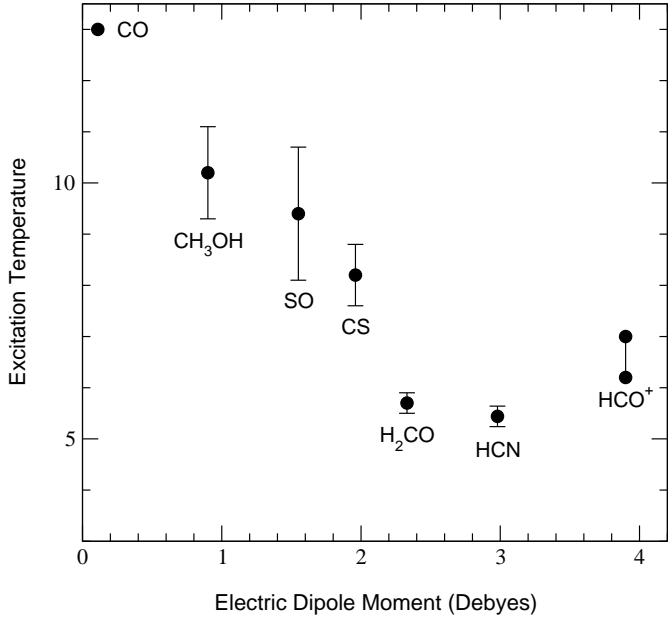}}
     \caption[]{
Excitation temperatures as a function of the dipole moment for the molecules
with two or more transitions detected towards the quiescent ambient cloud 
ahead of \hhd.  The $\Tex$ value used for CO, 13~K, is based on the assumption
that it is thermalized and equal to the kinetic temperature derived for the
\hco.
}
     \label{fmu} 
     \end{figure}
%

%
     \begin{table}
     \caption[]{Column Densities and Abundances}
     \label{tcdens}
     \[
     \begin{tabular}{lrrr}
     \noalign{\smallskip}
     \hline
     \noalign{\smallskip}   
\multicolumn{1}{c}{} &
\multicolumn{1}{c}{Column} &
\multicolumn{1}{c}{$X[$molecule$]/$} 
\\
\multicolumn{1}{l}{Molecule} &
\multicolumn{1}{c}{Density} &
\multicolumn{1}{c}{$X[$CO$]$} 
\\
     \noalign{\smallskip}
     \hline
CO$^{a}$& 3.8\N{17}  & 1.0        \\
\tco\   & 6.0\N{15}  & 1.6\N{-2}  \\
\cdo\   & 6.8\N{14}  & 2.8\N{-3}  \\
\hco\   & 1.0\N{13}  & 2.6\N{-4}  \\
\metha  & 6.4\N{13}  & 1.7\N{-4}  \\
H$_2$CO & 5.7\N{13}  & 1.7\N{-4}  \\
SO      & 3.2\N{13}  & 8.4\N{-5}  \\
SO$_2$  & 1.7\N{13}  & 4.4\N{-5}  \\
HCN     & 4.1\N{12}  & 1.1\N{-5}  \\
CN      & 3.4\N{12}  & 8.9\N{-6}  \\
CS      & 2.8\N{12}  & 7.3\N{-6}  \\
DCN     & 1.0\N{12}  & 2.6\N{-6}  \\
DCO$^+$ & 1.0\N{12}  & 2.6\N{-6}  \\
\\
HCS$^+$   & \mq3.6\N{11} & \mq9.4\N{-7}  \\
$c$-\cthd & \mq4.0\N{11} & \mq1.0\N{-6}  \\
HNCO      & \mq8.7\N{11} & \mq2.3\N{-6}  \\
CH$_3$CN  & \mq1.1\N{12} & \mq2.9\N{-6}  \\
NH$_2$CN  & \mq1.7\N{12} & \mq4.4\N{-5}  \\
HC$_3$N   & \mq2.6\N{12} & \mq6.8\N{-6}  \\
H$_2$CS   & \mq1.6\N{12} & \mq4.2\N{-6}  \\
HCOOH     & \mq3.3\N{12} & \mq8.6\N{-6}  \\
CH$_3$SH  & \mq3.4\N{12} & \mq8.9\N{-6}  \\
OCS       & \mq8.1\N{12} & \mq2.1\N{-5}  \\
NS        & \mq1.3\N{13} & \mq3.4\N{-5}  \\
H$_2$S    & \mq3.9\N{14} & \mq1.0\N{-3}  \\
     \noalign{\smallskip}
     \hline
     \end{tabular}
     \]
     \begin{list}{}{}
     \item[$^{a}$] Used a $^{12}$C$^{16}$O/$^{12}$C$^{18}$O ratio of 560.
     \end{list}
     \end{table}
%

\subsubsection{CS}

The excitation temperature derived from the two transitions is $\sim 8$~K. This
slightly higher excitation temperature than that derived from the other
molecules is possibly due to its slightly lower dipole moment.  As in the
previous cases, this excitation temperature was used to derive the column
density.

\subsubsection{SO$_2$}

Of the five SO$_2$ transitions observed, only two transitions were detected,
\JK{3}{1,3}{2}{0,2}\ and \JK{6}{0,6}{5}{1,5}.  If their emission is optically
thin, then their line ratio gives an unusually high excitation temperature,
$\sim 280$~K, much higher than that derived from other molecules. At this
temperature the \JK{10}{1,9}{9}{2,8}\ and \JK{10}{1,9}{10}{0,10}\ transitions
should have been detected with BIMA. Therefore, this temperature seems
unreliable.  If an excitation temperature of about 10~K is assumed, then the 
line ratio gives an \JK{3}{1,3}{2}{0,2}\ optical depth of 
10.  From the radiative transfer equation for the optically thick case, 
$\T{mb} = f \, (J[\Tex]-J[\T{bg}])$, the filling factor can be derived and, 
therefore, a rough estimation of the source size, $\theta$,
$f=\theta^2/(\theta^2+30^2)$ (30$''$ is the angular resolution of the spectra).
The peak intensity of the \JK{3}{1,3}{2}{0,2}\ line, $\sim 0.2$~K, would imply
a source size of only 5$''$ if the line is optically thick.  However, the BIMA
maps of this transition show that the emission arises from a significantly
larger area, $\sim 19''$ (Paper II). Thus, we suggest that the
\JK{6}{0,6}{5}{1,5}\ line emission is ``anomalous'', although we cannot discern
if this is an instrumental effect or if it is a real anomaly. In order to
estimate the SO$_2$ column density we used the \JK{3}{1,3}{2}{0,2}\ transition,
assuming an excitation temperature of 10~K.

\subsubsection{Other molecules}

The CO and its isotopes have a low dipole moment, 0.11 Debyes, so their
emission is likely thermalized.  Therefore, the excitation temperature used to
derived the CO column density is 13~K.
Figure~\ref{fmu} shows the values of the excitation temperature {\em
versus} the dipole moment of the analyzed aforementioned molecules. With the
exception of the \hco, there is a trend of decreasing $\Tex$ with increasing
dipole moment, $\mu$, which is expected since higher $\mu$ implies more severe
subthermalization.  Fig.~\ref{fmu} can be used to roughly extrapolate
the excitation temperature for the other observed molecules.  Thus, we adopted
a $\Tex = 10$~K for those molecules with $\mu \la 2$~Debyes (CN, OCS, H$_2$S,
HCS$^+$, H$_2$CS, HNCO, HCOOH and CH$_3$SH) and $\Tex = 6$~K for the rest of
the molecules ($c$-\cthd, HC$_3$N, CH$_3$CN and NH$_2$CN), which indeed have a
dipole moment higher than 3 Debyes. 

Table~\ref{tcdens} shows the beam-averaged  column densities derived from the
observations.  The fractional abundances of the observed molecules with respect
to the CO molecule are also listed in this table (note that the estimated
fractional abundance relative to the CO is a beam-averaged value).

\section{Discussion}

\subsection{Chemical Comparison with other dense molecular regions}

%
     \begin{figure} 
     \resizebox{\hsize}{!}{\includegraphics{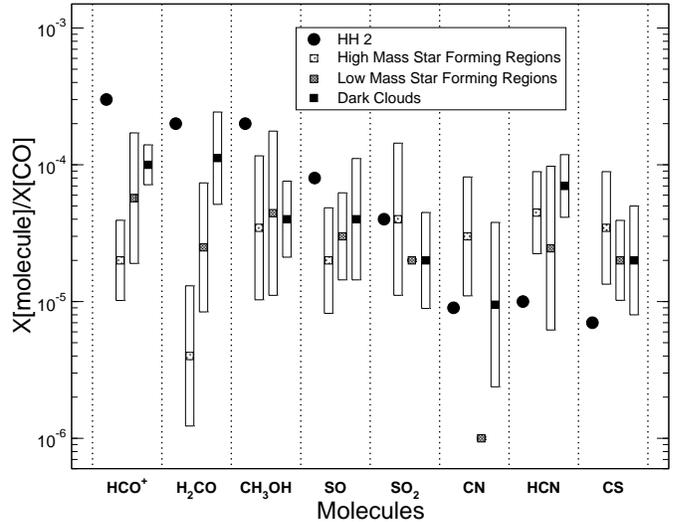}}
     \caption[]{
Relative molecular abundances (with respect to the CO) for \hhd\ (filled
circles), high mass star forming cores,  low mass star forming cores and dark
clouds (\ie\ without star formation activity).  The filled square within the 
bar and the bar show the logarithmic median and the logarithmic standard 
deviation, respectively, of the sample.  
The sample for the high mass star forming clouds is: 
W3 (Helmich \& van Dishoeck \cite{Helmich97}),
M17 and Cepheus A (Bergin \et\ \cite{Bergin97}). 
The sample for the low mass star forming molecular clouds is: 
IRAS~16293 (van Dishoeck \et\ \cite{Dishoeck95}), NGC~1333~IRAS4A (Blake \et\
\cite{Blake95}),the Serpens S68 (McMullin \et\ \cite{McMullin00}), 05338-0624
(McMullin, Mundy \& Blake \cite{McMullin94}) and several other regions for 
\form\ (Dickens \& Irvine \cite{Dickens99}) and \metha\ (Kalenskii \& Sobolev
\cite{Kalenskii94}).  
The sample for the quiescent dark molecular clouds is: 
OMC-1N (Ungerechts \et\ \cite{Ungerechts97}), 
TMC-1 (Pratap \et\ \cite{Pratap97}), 
L134N (Dickens \et\ \cite{Dickens00}), and 
the CB clouds for \form\ (Turner \cite{Turner94}).
The references of several of the selected molecular clouds give the relative
chemical abundances in different positions of the cloud, so most of the data 
plotted in this figure were obtained averaging four or more points (the SO$_2$
and CN values in low mass star forming cores are from only one region). 
}
     \label{fxmol} 
     \end{figure}
%

In order to be able to establish the effects of the shock-induced radiation
from \hhd\ on the chemical composition of the quiescent clump ahead of the HH
object, it is necessary to compare the relative abundances of the molecules
detected with respect to other environments associated with dense molecular
material. Before the radiation reached the clump, this molecular cloud had
likely the properties of the starless dark molecular cores.  It is well known 
that there are important chemical differences between starless dark clouds and
star forming regions, and that there are important gradients within  the same
molecular cloud, a consequence of the evolutionary effects or due to the
presence of nearby star formation sites (\eg\ van Dishoeck \& Blake
\cite{Dishoeck98}).  Since we want to compare the general chemical properties
of the molecular clouds with those from the molecular emission ahead of \hhd,
we have estimated the logarithmic median and standard deviation of the relative
abundance (with respect to CO) of the molecules detected in \hhd\ and compare 
this with a sample of three different types of molecular environments: high 
mass and low mass star forming molecular clouds, and starless dark molecular
clouds.  Figure~\ref{fxmol} shows shows this comparison.

Interestingly, from Figure~\ref{fxmol}, the observed molecules ahead of \hhd\
can be classified into four groups, depending on their abundance with respect
to other molecular clouds. 
(1) Strongly enhanced molecules: \hco, \metha\ and \form. 
Their relative
abundances are enhanced mostly within a factor of 5 to 10 with respect to the
typical values found in the molecular clouds.  In particular, the enhancement 
of \hco\ and \form\ in \hhd\ with respect to the high mass star forming regions
is more than an order of magnitude, but only by a factor of 2-3 with respect to
the dark molecular clouds. 
(2) Weakly enhanced molecules, SO and SO$_2$.  Their enhancement with respect 
to the molecular clouds is a factor of 2-4.  The only exception is that the
SO$_2$ has similar relative abundances as the massive star forming regions. 
(3) The CN shows no apparent enhancement with respect to the dark molecular
clouds, whereas it is depleted with respect to the high mass star forming
regions. 
The strong enhancement (an order of magnitude) with respect to the low mass
star forming regions may not be significant since there is only one data point. 
(4) Depleted molecules: CS and HCN. In \hhd\ these molecules are clearly
depleted with respect to the other molecular clouds, with depletion factors in
the 3 to 7 range.

\subsection{Comparison with the VW99 Models}

Here we briefly compare our observational results with the abundances derived
in the VW99 models. A more detailed model of the HH2 clump (in light of the
present observations) will be the subject of a future paper. Columns 3 to 6 of
Table~\ref{tmodel} lists the best matched theoretical column densities and the
visual extinction at which this column density was obtained for two models out
of the grid from VW99 for clump densities of 1\N{4}\cmt\ (Model 4) and of
1\N{5}\cmt\ (Model 5), 3 years after irradiation of the clump has started.

Table~\ref{tmodel} clearly confirms the qualitative results from VW99 where 
many
species were predicted to be abundant in clumps ahead of HH objects. The VW99
models were not specific to any particular HH object but use an extensive gas
and grain phase chemistry to follow the formation of a clump ahead of a shock
with a fixed enhanced radiation field (equivalent to 20 times the ambient
interstellar radiation field) acting upon it.  The clump is treated as a
one-dimensional slab extending up to 6 visual magnitudes to the center.

%
     \begin{table}
     \caption{}
     \label{tmodel}
     \begin{tabular}{lrrccrr}
     \hline
\multicolumn{1}{l}{}&
\multicolumn{1}{c}{HH 2}    &
\multicolumn{2}{c}{Model 4} &&
\multicolumn{2}{c}{Model 5} 
\\
\cline{3-4} \cline{6-7}
\multicolumn{1}{l}{Molecule}&
\multicolumn{1}{c}{$N_{\rm molecule}$} &
\multicolumn{1}{c}{$N_{\rm molecule}$} &
\multicolumn{1}{c}{$A_{\rm V}$} &&
\multicolumn{1}{c}{$N_{\rm molecule}$} &
\multicolumn{1}{c}{$A_{\rm V}$} 
\\
\hline
CO       & 3.5\N{17} & 3.1\N{17} &3.2 && 3.2\N{17} &3.1 \\
HCO$^+$  & 1.0\N{14} &   3\N{13} &1.0 && 2.5\N{12} &1.1 \\
CH$_3$OH & 6.4\N{13} & 7.8\N{13} &1.3 && 4.0\N{14} &1.1 \\
H$_2$CO  & 5.7\N{13} & 3.3\N{13} &6.0 && 5.7\N{13} &2.0 \\
SO       & 3.2\N{13} & 3.0\N{13} &3.4 && 5.7\N{13} &2.8 \\
SO$_2$   & 1.7\N{13} & 1.8\N{13} &4.1 && 1.9\N{13} &3.8 \\
HCN      & 4.1\N{12} & 4-5\N{12} &1-2 && 5.0\N{12} &1.3 \\
CN       & 3.4\N{12} &   4\N{12} &1.7 && 3.4\N{12} &1.3 \\
CS       & 2.8\N{12} &   8\N{12} &1.6 && 5.0\N{12} &1.9 \\
     \hline
     \end{tabular}
     \end{table}
%

In general, the VW99 models suggest that: (i) HH2 is a young object,  probably
on the order of 1000 years;  (ii) the best estimate for the volume density of
the clump is $<$ 10$^5$ cm$^{-3}$, somewhat less than the density derived from
observations; (iii) the visual extinction at the center of the SO peak emission
with respect to \hhd\ is about 5-7 mags; VW99 models only extend to 6 mag;
although quantitatively the models do not match the observed abundances for
every species at the maximum $A_{\rm V}$, most of them are well represented at
some visual extinction. In fact, Model 4 of VW99  reproduces the observed
abundance (within half order of magnitude) for every species for an optimal
visual extinction, different from species to species: this may be an indication
of the displacement found with the BIMA array (Paper II). More specifically, it
is interesting to note that HCO$^+$, the best tracer of HH clumps, is
underabundant for visual extinction larger than $\sim$ 1 mag. This is because
at high visual extinction the clump is more shielded from the radiation and
this would limit the amount of ionized carbon, yielding a low production of
HCO$^+$; a higher radiation field may well therefore increase its abundance. 
The presence  of X-ray emission in \hhd\ (Pravdo \et\ \cite{Pravdo01}) may have
some  effects on the chemistry.  A detailed model for \hhd\ which takes such 
considerations into account is in preparation.

\section{Conclusions}

We presented a 3--0.8~mm molecular line survey toward the quiescent molecular
clump ahead of the bright \hhd\ object.  A total of 14 species were detected
(including different isotopes and deuterated species) and the upper limits of
12 more species were obtained. Multi-transitions observations of \hco, SO and
\metha\ show that the gas is not only quiescent, but also cold, $\sim 13$~K,
in spite of being close to \hhd.  The density of the cloud is roughly
3\N{5}\cmt.  Comparisons of the relative abundances (with respect to CO)
were made with different dense molecular gas environments, including quiescent,
starless dark clouds, and active low and high mass star forming cores. We
confirm the peculiar chemical composition of the gas ahead of the HH objects, a
consequence of the mantle removal and further photochemistry produced by the
strong radiation generated in the shocks. In particular, we found that:
\begin{itemize} 
\item \hco, \metha\ and \form\ are strongly enhanced with respect to other
molecular cores. 
\item SO and SO$_2$ are weakly enhanced with respect to other molecular
environments.  The only exception is that the SO$_2$ has similar relative 
abundances as the massive star forming regions.
\item CN does not show a significant enhancement or depletion with respect to 
the starless dark clouds. However it is underabundant with respect to high 
mass star forming regions. 
\item HCN and CS show clear depletion when the \hhd\ relative abundances are 
compared with most of the molecular clouds.  
\end{itemize}

Finally, the chemical composition of \hhd\ confirms the qualitative results of
the VW99 complex chemical model that follows the chemical behavior of a
molecular clump irradiated by a HH object.

\begin{acknowledgements}

We thank S. Curiel for providing the [SII] images of \hhd. We thank W. Dent for
his valuable comments.  JMG acknowledge
support by NSF grant AST-99-81363 and by RED-2000 from the Generalitat de
Catalunya.  RE and JMG are partially supported by DGICYT grant PB98-0670
(Spain). SV and DAW thank  PPARC for supporting their research.

\end{acknowledgements}

\begin{appendix}
\section{The Population Diagram Analysis\label{Apendix}}

The population diagram (also called rotation diagram in the literature)
analysis has been extensively used in the literature to obtain the temperature
and the column densities for molecules with several transitions observed (\eg\ 
Linke, Frerking \& Thaddeus \cite{Linke79}; Turner~\cite{Turner91}).  In the
use of the population diagram, we assume LTE conditions (\ie\ all the
transitions are populated with a single excitation temperature $\Tex$),
optically thin emission and $\Tex \gg T_{bg}$. Under these assumptions the
population diagram can be described with the following expression:
\begin{equation}\label{drot}
{\rm log} \, I =
{\rm log} \, \frac{N_{\rm ba}}{Q_{\rm rot}} - \frac{{\rm log} \, e}{T} E_u
\end{equation}
with
\begin{equation}
 I = \frac{3 k \int T_{\rm mb} dv}{8 \pi^3 g_{\rm K} g_{\rm I} S \mu^2 \nu}
\end{equation}
where $g_{\rm K}$ and $g_{\rm I}$ are the $K$-level and nuclear spin
degeneracies respectively,  $\mu$ the electric dipole moment of the molecule,
S, $\nu$, $ T_{\rm mb}$ and $E_u$ are the line strength, frequency,  main-beam
brightness temperature and the energy of the upper state (in temperature
units) of the transition, $N_{\rm ba}$ the total beam-averaged column density
of the molecule, $Q_{\rm rot}$ the rotational partition function of the
molecule.  This formula appears in the population diagram (in which the left
hand part of the formula is plotted {\it versus} $E_u$) as a straight line.
However, several factors can yield temperature and column densities that are
significantly different from their true values: subthermalization (or non-LTE
excitation), significant optical depths, excitation gradients within the beam
and beam dilution (when different angular resolutions are used).  Goldsmith \&
Langer  (\cite{Goldsmith99}) describe the effects that subthermalization and
non-null optical depths have on the population diagram.

\begin{figure} 
     \resizebox{\hsize}{!}{\includegraphics{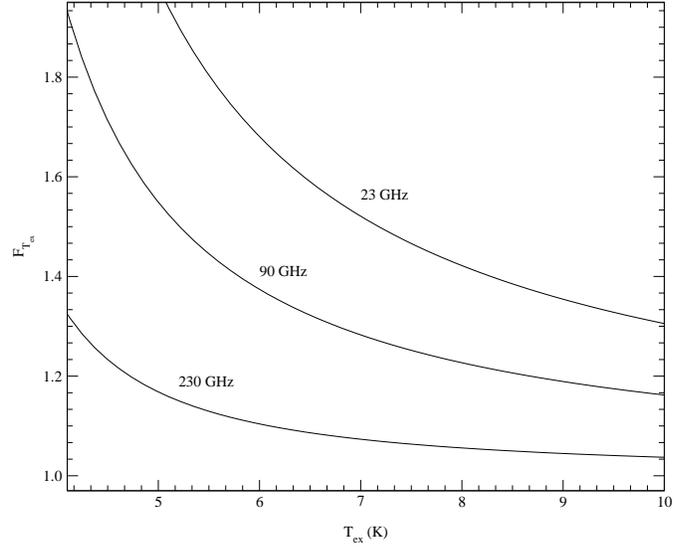}}
     \caption[]{
$F_T$ as a function of temperature for three different frequencies.
}
     \label{ftex} 
     \end{figure}

In order to better use the population diagram technique we just assume that all
the rotational levels are populated with the same excitation temperature (which
in most cases is likely lower than the kinetic temperature: see Goldsmith \&
Langer \cite{Goldsmith99}).  In this case the expression is:
\begin{equation}\label{drotb}
{\rm log} \, I =
{\rm log} \, \frac{N_{\rm ba}}{Q_{\rm rot}} - \frac{{\rm log} \, e}{T} E_u -
{\rm log} \, C_{\tau} -
{\rm log} \, F_T
\end{equation}
with $C_{\tau}$ and $F_T$ are defined as:
\begin{equation}
C_{\tau} = \frac{\tau}{1-e^{-\tau}}
\end{equation}
and
\begin{equation}
F_T = \frac{J_{\nu}[\Tex]}{J_{\nu}[\Tex]-J_{\nu}[\T{bg}]}
\end{equation}
$J_{\nu}[T]$  is the Planck function in temperature units.  Note that when 
$\tau \ll 1$ and $\Tex \gg T_{bg}$, equation~\ref{drotb} becomes
equation~\ref{drot}.  However, for those regions where the temperature is low
and with moderate densities, $F_T$ should be taken into account, especially for
the low frequency transitions (see Fig.~\ref{ftex}).  $C_{\tau}$ should also be
taken into account even for moderately small opacities: \eg\ $C_{\tau}=1.25$
for $\tau = 0.5$.

%
%

The optical depth for each transition can be estimated if true column
density, $N_{\rm true}$, is known.  Since $N_{\rm true} \times f = N_{\rm ba}$,
the optical depth of a given rotational transition is given by:
\begin{equation}
\tau _{J,J+1} = \frac{8 \pi^3 \mu^2}{3 h} \frac{N_{\rm ba}}{f} 
\frac{J e^{-aJ(J+1)} (e^{2aJ}-1)}{Q_{\rm rot}}
\end{equation}
Therefore, if the filling factor $f$ is included as an additional free
parameter (with $T$ and $N_{\rm ba}$), the optical depth can be corrected
whenever using the population diagram method.

To estimate the best set of solutions we calculated the $\chi^2$ value for each
set of parameters, $T_{\rm rot}$, $N_{\rm ba}$ and $f$, taking into
account the uncertainties (rms noise and calibration errors) in a similar way
as that described by  Nummelin \et\ (\cite{Nummelin98}) and Gibb \et\
(\cite{Gibb00}): 
\begin{equation}
\chi^2 = \sum_{i=1}^n  \left (
\frac{I^{\rm obs}_{i}-I^{\rm calc}_{i}}{\sigma^{\rm obs}_{i}} \right )^2 
\end{equation} 
For those transitions undetected, the 3-$\sigma$ upper limit for each 
transition was used whenever $I^{\rm calc}_{i} \le 3\times \sigma$. The
1-$\sigma$ uncertainties for $T_{\rm rot}$ and $N_{\rm ba}$ were obtained
from the 68\% confidence region of $\chi^2$ (\eg\ Gibb \et\ \cite{Gibb00}),
which is enclosed within the $\chi^2_{\rm min} + \Delta \chi^2$ interval, with
$\Delta \chi^2=x/(N-P)$ where $N$ is the number of data points (including those
undetected transitions that satisfy the aforementioned condition), $P$ is the
number of free parameters (3 when the opacity correction is taken into account
and 2 when optically thin emission is assumed), and $x$ equals to 2.3 and 3.5
for 2 and 3 free parameters, respectively (Lampton, Margon \& Bowyer
\cite{Lampton76}).

\end{appendix}


\end{document}